\documentclass[a4paper,secnumarabic,amssymb, nobibnotes, aps, prd,10pt,floatfix]{revtex4-1}
\usepackage{graphicx}
\usepackage{epstopdf}
\usepackage{verbatim}
\usepackage{caption}
\usepackage{subcaption}
\usepackage{color}

\begin{document}

\title{Self-similar Charge Transport in Gapped Graphene}

\author{D\'iaz-Guerrero D. S.\textsuperscript{1}*, Gaggero-Sager L. M.\textsuperscript{1}, Rodr\'iguez Vargas I.\textsuperscript{2}, \& Naumis G.G.\textsuperscript{3}}
\affiliation{1.Facultad de Ciencias, Universidad Aut\'onoma del Estado de Morelos. Cuernavaca Morelos, M\'exico. 
2. Unidad de F\'isica, Universidad Aut\'onoma de Zacatecas. Zacatecas Zacatecas, M\'exico. 
3. Instituto F\'isica, Depto. de F\'isica-Qu\'imica, Universidad Nacional Aut\'onoma de M\'exico (UNAM). Apdo. Postal 20-364, 01000, M\'exico D.F., M\'exico.}
\begin{abstract}
 A new type of self-similar  potential is used to study a multibarrier system made of graphene. 
 Such potential is based on the traditional middle third Cantor set rule combined with a scaling of the barriers height.
 The resulting transmission coefficient 
 for charge carriers, obtained using the quantum relativistic Dirac equation, shows a surprising 
 self-similar structure. The same potential does not lead to a self-similar transmission when applied 
 to the typical semiconductors described by the non-relativistic Schr\"odinger equation.
 The proposed system is one of the few examples in which a self-similar structure produces the 
 same pattern in a physical property. 
 The resulting scaling properties are investigated as a function of three parameters: the height of the main barrier, 
 the total length of the system and the generation number of the potential. These scaling properties are first identified individually and then 
combined to find general analytic scaling expressions.\newline
\newline
* diazd@uaem.mx
\end{abstract}

\maketitle
Graphene\cite{novoselov2004} is considered as one of the most promising new materials. 
This first truly two-dimensional crystal \cite{novoselov2005} has impressive physical properties 
\cite{novoselov2004,balandin2008}. As a consequence, graphene is not only important from a technological 
point of view, but 
also is considered as an inflexion point in quantum physics. For instance, charge carriers in graphene follow an effective quantum 
relativistic (Dirac) equation instead of the usual Schr\"odinger one. This leads to new effects \cite{Geim2007,Novoselov2011,Stone2012}. Specially, 
the scattering produced by external potentials or impurities is different from what is observed in ordinary materials. Charge 
carriers can be transmitted perfectly due to the Klein effect and random sustitutional impurities can produce multifractal 
states \cite{Naumis1994,Naumis2002}. Thus, a whole new world is open to investigate the possible effects of external potential geometries, which
can be imposed to graphene by different means, as for example, using certain type of substrates, strain, electrostatic gates, impurities, electromagnetic 
fields, etc. \cite{Giovannetti2007,CastroNetoReview2009}.  \newline

For applications, an important issue is how to engineer substrate-induced bandgap opening in epitaxial graphene \cite{Giovannetti2007}. This leads to investigate how different multibarrier geometries affect the properties of  graphene. Among these possible geometries, 
self-similarity has a paramount importance, since scale invariance is a fundamental property of many natural phenomena, as can be corroborated in reports that ranges from 
the distribution and abundance of species, temporal occurrence of earthquakes, and even in the growth of complex networks and trees \cite{JHarte1999Science,ACorral2004PRL,CSong2006NaturePhysics,CEloy2011PRL}. 

From the technological standpoint, self-similarity can also be exploited to
produce useful devices \cite{MSun2006PRB,FMiyamaru2009APL,HXDing2012APL} .  Particularly, photoconductive fractal antennas show an efficient multiband emission of terahertz radiation owing to the self-similarity of the fractal structure \cite{FMiyamaru2009APL}. 

Despite that self-similar structures are getting plenty of attention \cite{EMacia2006Review}, only a handful of experiments corroborate 
a self-similar behavior in its physical properties \cite{ALavrinenko2002PRE,BHou2004APL,JEstevez2012JAP}. 
Likewise, many theoretical works  claim that physical properties, such as tranmission or 
reflection probabilities, display self-similarity. Unfortunately, most of them are just a matter of visual perception, 
since the scaling properties are never reported. Even in the case of quasicrystals, which 
are considered as one  of the archetypical examples of self-similar structures, it has been elusive to find 
the scaling properties of the corresponding physical properties \cite{Naumis-Aragon1996,Nava-Taguena-delRio2009}.

Here we propose a novel self-similar multibarrier structure in graphene in which the main physical 
property, namely charge transport, presents a surprising self-similar pattern. This
self-similar structure does not produce a similar behavior in a usual semiconducting material described 
by the Schr\"odinger equation. Thus, two essential ingredients are shown to be into play:
a self-similar structure and quantum relativistic mechanics.

\begin{figure*}[ht]
\includegraphics[width=4in, height=4in]{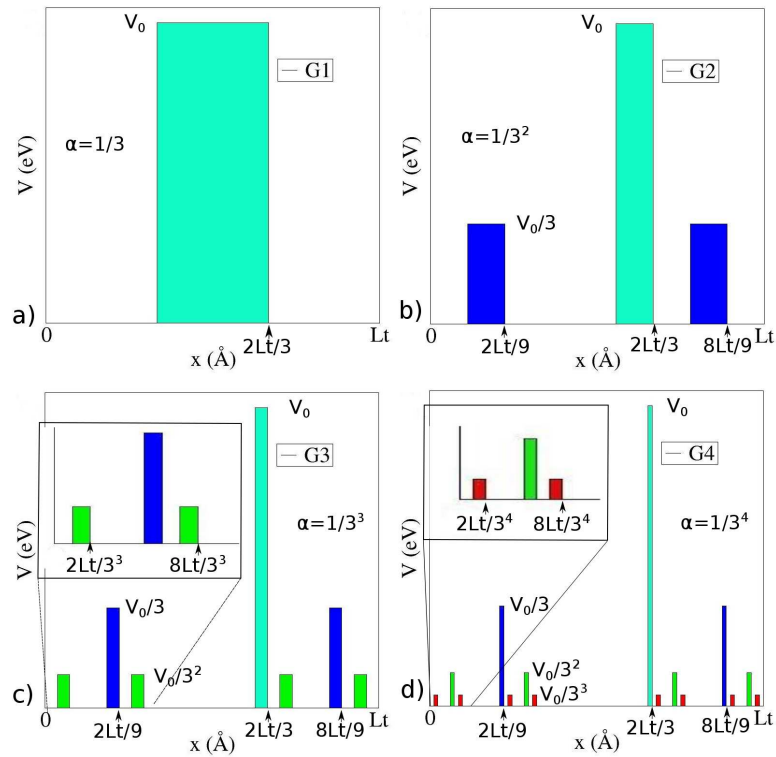}
\caption{\label{potential} \textbf{Potential construction}. First four generations (G1,G2,G3,G4) of the 
multibarrier potential. Each barrier is scaled (we denote the scaling factor by $\alpha$) in its width, and so
the barrier width is $Lt/\alpha$. The barriers added in the last iteration are scaled in its height. For 
example, the height of the green ones are one third of the height of the blue ones. The zooms in G3 and G4 illustrate
how the potential resembles a previous generation after an appropiate rescaling.}
 \end{figure*}

To build our multibarrrier structure in one direction (the $x$ axis),  we implement a 
variant of the middle third Cantor case \cite{cantorSet} (see supplementary material) using square barriers as indicated in Fig. \ref{potential}. We begin with a line-segment of length $Lt$, we divide it in thirds
and put a rectangular barrier of energy $V_0$ in the middle third (its width is $Lt/3$). We call this the generation one ($G1$). 
The second step  consists in taking the remaining 
line-segments, divide them into thirds and place within scaled copies of the existing barrier. We scale the height and width of the barrier 
by a scaling factor of $1/3$. Additionally the existing barrier is also scaled in its width. Since scaling is a contraction transformation, 
each scaled copy will have a unique fixed point, \textit{see the fixed point theorem in metric spaces} \cite{ApostolBook}. By fixed point we meant an element of the function's domain that is mapped to itself by the function. In this case, this point 
is a fixed side of the barriers, which  will be chosen as the right-hand side. So, every successive iteration (generation) ($Gj$) of the structure 
is obtained by repeating the second step of the construction for the remaining line-segmentes, using as the ``existing barrier'' those added in 
the last iteration, as shown in Fig. \ref{potential}.\newline

This potential can be described as follows. The $j$-th generation of the potential will consists of $N_{Gj}$ barriers; the corresponding 
barriers (those added in the $j$-th iteration) 
has a width $W_G(j)$ and energy (height) $V_G{j}$. Such barrier is obtained as a scaled replica of barriers from generation $G(j-1)$,
\begin{eqnarray}
N_{Gj}=2^{Gj}-1, \nonumber \\
W_{Gj}=\alpha=\frac{Lt}{3^{Gj}}, \nonumber \\
V_{Gj}=\frac{V_0}{3^{Gj-1}}. \nonumber
\end{eqnarray}

This bidimensional potential does not depend on the perpendicular direction ($y$ axis), i.e.,  $V(x,y)=V(x)$ and can be adapted to graphene  
\cite{novoselov2004,KNovoselov2005Nature,YZhang2005Nature,Beenakker2009} by means of 
symmetry breaking substrates \cite{SZhou2007NMaterials,JVGomes2008JPCM,VHNguyen2011SST}, by gated graphene or in other
systems, like optical analogies, with a linear spectrum \cite{cheianov2007,Hartmann2010,Hartmann2014}.

Let us now study the transmission coefficient of charge carriers in this multibarrier system. For graphene, we use 
the Viana-Gomes et al. formalism \cite{JVGomes2008JPCM} to compute the transmission coefficient of electrons
in gapped graphene. The Hamiltonian for the Dirac electrons under a potential $V(x)$ is given by 
\begin{equation}
H=v_F{\bf \sigma}\cdotp {\bf p} + \sigma_z V(x),
\end{equation}
where $V(x)=m(x)v_F^2$ if $x\in[x_1,x_2]$ and zero in other case. $V(x)$ defines the region where there is
a mass term ($q$-region) and where carriers are massles Dirac fermions ($k$-region).
The first term of the Hamiltonian is just the Dirac equation used to describe electrons in graphene,
where $v_F$ is the Fermi velocity, ${\bf \sigma}$ the set of Pauli matrices and ${\bf p}$ the moment.
For normal incidence on the $x$ direction the corresponding wavefunctions are \cite{JVGomes2008JPCM}
\begin{equation}
\psi_k^{\pm}(x)=\frac{1}{\sqrt{2}}\left(
\begin{array}{c}
 1 \\
 u_\pm
 \end{array}\right)
 e^{\pm ik_xx},
\end{equation}
and
\begin{equation}
\psi_q^{\pm}(x)=\frac{1}{\sqrt{2}}\left(
\begin{array}{c}
 1 \\
 v_\pm
 \end{array}\right)
 e^{\pm iq_xx},
\end{equation}
where $u_\pm =\pm \mbox{sign}(E)$ and $v_\pm =\frac{E-V(x)}{\hbar v_F(\pm q_x-ik_y)}$. Using these equations, one establishes the continuity conditions 
when the wave goes from the $k$-region to the $q$-region.\newline
\begin{figure*}[ht]
        \includegraphics[width=4.5in, height=4.5in]{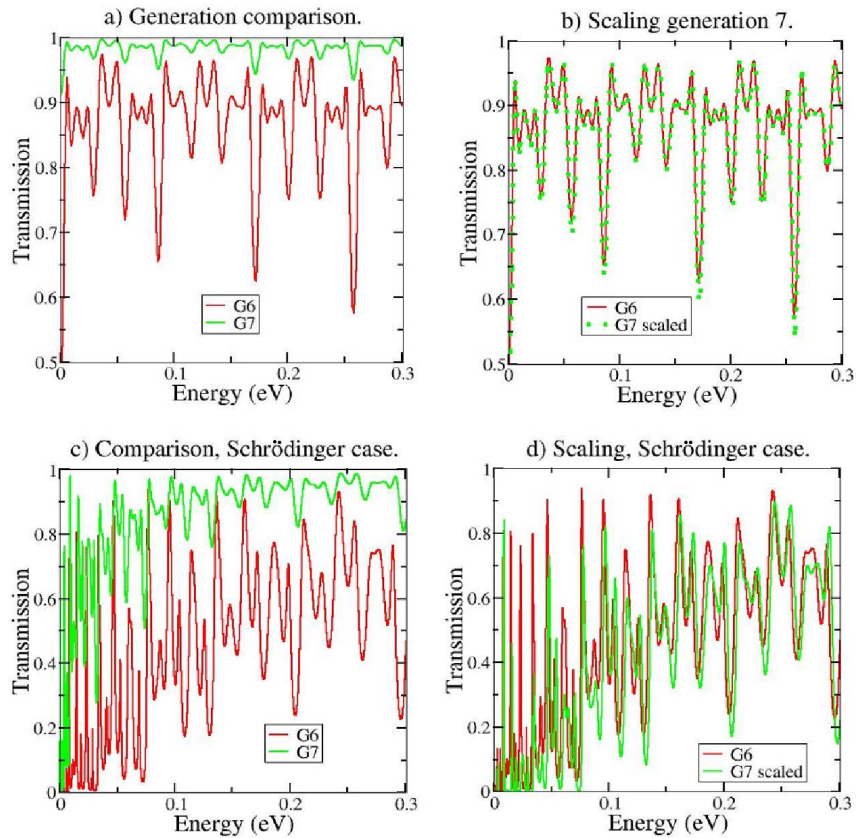}
        \caption{\label{gen-comparison} \textbf{Transmission coefficient scaling for different generations}. a) Transmission as a function of energy 
        for the sixth ($G=6$) and seventh ($G=7$) generation of multibarrier graphene using the 
        parameters $V_0=1.0$ eV and $Lt=3500 \ \mathring{A}$. b) Scaled
        transmission for $G=6$ and $G=7$. The scaling transformation is given by the power law Eq. (\ref{g-scaling}), 
        which gives an excellent agreement of the curves. In c) and d), the transmission and scaled transmission 
        respectively are presented
        for the same potential but for a typical semiconductor which follows the Schr\"odinger equation. No self-similar scaling
         is observed.}
\end{figure*}

With the potential described in Fig. \ref{potential} at hand, we compute the transmission coefficient, defined as
\begin{equation}
 T_G(E)=\frac{|\psi_k^{\pm}(0)|^2}{|\psi_k^{\pm}(Lt)|^2}
\end{equation}

using the transfer matrix  formalism \cite{SokoulisBook,JVGomes2008JPCM,IRVargas2012JAP}. 
The transfer matrix can be obtained by imposing the continuity conditions of 
the wave functions in each region of the self-similar structure.
From this we can extract the transmission coefficient for a given energy, see for instance \cite{JVGomes2008JPCM, IRVargas2012JAP}.

Since the construction of the multibarrier structure is based on the length-scaling of the previous generation barriers,
and the addition of energy-scaled
new barriers, we search for relations between transmission curves corresponding to different generations of 
the multibarrier structure. So, we fix the total length of the structure, denoted by $Lt$ (in this case we use the value of 
$Lt=3500 \ \mathring{A}$) and the height of the main barrier, denoted by $V_0$ (using the value of $V_0=1$ eV).\newline

In figure \ref{gen-comparison}a we show the transmission curves corresponding to generations 6 and 7 of the potential, denoted by 
$T_6(E)$ and $T_7(E)$ respectively. As we can see, it is qualitatively clear that they are related by some kind of transformation. 
This transformation turns out to be $T_6(E)=\left[T_7(E)\right]^9$,
where the subindex stands for the generation number, as seen in figure \ref{gen-comparison}b. Thus, $T_6(E)$ are $T_7(E)$ scaled by
a single exponent. As may be expected, other bigger generations also follow the same rule. More formaly
\begin{equation}
T_G(E)\approx \left[ T_{G+1}(E)\right]^9,
\label{g-scaling}
\end{equation}
where $G\geq 6$ and $G$ stands for the generation number. This means that any two transmission curves that differ only in the generation of their 
correspondig potential, and are related by some power of the form $9^m$, where $m$ is the difference between the generations. 
This suggests that the transmission  curves follows an scaling rule which is a signature of the self-similar 
phenomena. In fact, here we are in presence of the first true self-similar physical property of graphene. 
Morever, even for perfect fractals it is known that the physical properties usually follow a multifractal 
behavior \cite{Nava-Taguena-delRio2009,NaumisMultiFrac2007}
but here only one single fractal exponent is obtained.\newline

It is important to remark here that there are always  finite size effects due to the break in the scaling of the potential, which affects the first generations. 
In this particular case, one cannot extend self-similarity beyond the length of the system. This effect is explained in the supplementary section where we present 
the transmittance from generation $G1$ to $G9$. For low generations, the transmittance is basically dominated by the biggest energy barrier. Between $G5$ and $G6$, 
there is a dramatic change in the behavior since now the system has transmittance for energies lower than the main barrier height. As generations increase, 
scaling appears. Generation $6$ is the crossover since as shown in the supplementary section, geometrical finite size effects in the potential die out after it. Also, 
is clear that in real systems there is also another  finite size effect at the bottom of the hierarchy, where the Dirac approach can not be made due to the rapid variation 
of the potential at atomic distances.

To highlight the difference with a typical Schr\"odinger semiconductor, in figure \ref{gen-comparison}, panels  c) and d), we
present the transmission calculated for the same potential generations using Al$_{x}$Ga$_{1-x}$As/GaAs. In such system, the height of
the quantum wells are controlled by the Aluminium concentration ($x$).

So far, we have found a scaling between transmission curves that correspond to different generations of the potential, 
but it seems 
possible to extend the search for scaling features to other parameters. 
We have two options at hand, the energy of the starting 
barrier and the total length of the structure.\newline
\begin{figure*}[ht]

\includegraphics[width=4.5in, height=4.5in]{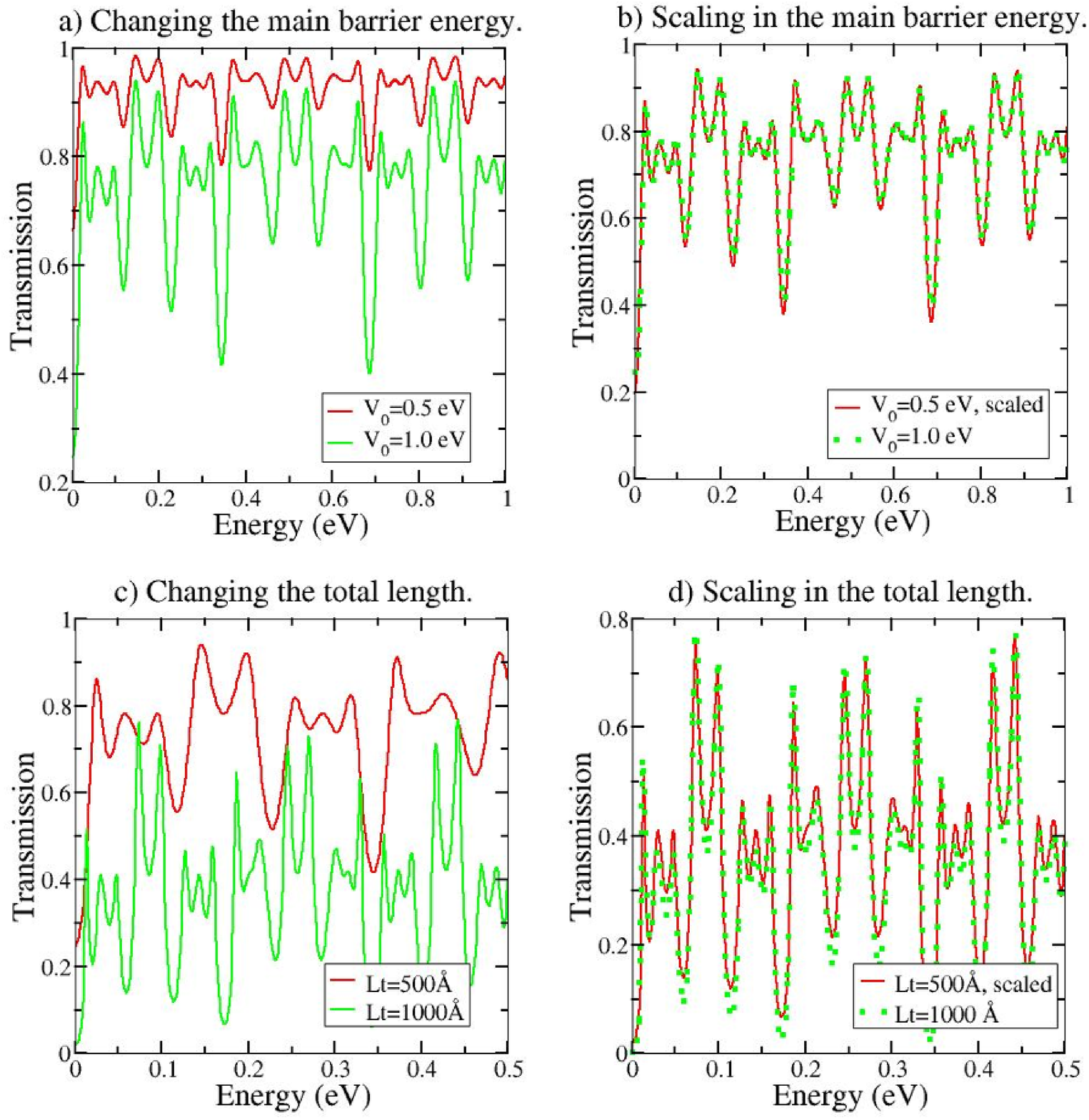}
\caption{\label{H0Lt-comparison}\textbf{Scaling between barrier energy}. a) Comparison between transmission curves for different values of $V_0$ 
using generation 6.  This case corresponds to  the transmission for two potentials with different heights 
of the main barrier, in this case $V_0=0.5$ eV and $V_0=1.0$ eV. All other parameters are equal.
b) Comparison between the scaled plots of a). The scaling consists in raising $V_0=0.5$ eV to the fourth power.
\textbf{Scaling between lengths}. 
c) Comparison between the transmission curves for different lengths ($Lt$). In this case it is quite clear
that the scaling cannot be just raising one curve to a certain power.
d) Scaling of the curve with length $Lt=500 \ \mathring{A}$. This scaling is made of two parts, raising the
curve to the fourth power and multiplying its energy by two.}
\end{figure*}

Let us explore now the first option, that is, we take a fixed generation of the potential, say generation 6, and make the calculation 
for two starting energies. Lets take $V_0=0.5$ eV and $V_0=1.0$ eV. In this case, the appearance of the curves seen in 
figure \ref{H0Lt-comparison}a,
resembles those corresponding to a change in the generation of the potential.
In the case of the nonrelativistic transmission coefficient (e. g. GaAs), for the same geometric set up, one observes in general a 
shift of the first transmission resonance. Nevertheless, in this case the general behavior of the 
transmission curve is qualitatively the same. Due to the resemblance of these curves with the ones of changing generations, we try the same kind of transformation but with 
a different power. In this case, one has to raise the curve corresponding to $V_0=0.5$ eV to the fourth power 
to get the one corresponding to $V_0=1.0$ eV, see figure \ref{H0Lt-comparison}b. Thus $T_{1.0}(E)\approx T_{0.5}(E)^4$,
where the subindex stands for the $V_0$ of the respective structure. In fact, this scaling behavior is more general, i. e.,
it applies for every multiple of two, resulting in the expression
\begin{equation}
T_{\frac{1}{k}V_0}(E)^{k^2}\approx T_{V_0}(E).
\label{v-scaling}
\end{equation}

Finally we compare curves corresponding to different lengths of the whole structure, denoted by $Lt$. Although 
nothing suggests scaling in this variant at all, see figure \ref{H0Lt-comparison}c, our previous results lead us to a 
different conclusion. From the results for the case of changing $V_0$, we try to find a similar scaling law for $Lt$, but 
in this case, the scaling in the energy gives the relation,
\begin{equation}
T_{\frac{1}{\alpha}Lt}(\frac{1}{\alpha}E)^{\alpha^2}\approx T_{Lt}(E).
\label{Lt-scaling}
\end{equation}
With this scaling rule applied to the curve corresponding to $500$ $\mathring{A}$, a very 
good approximation is obtained, see figure \ref{H0Lt-comparison}d.

Combining expressions (\ref{g-scaling}), (\ref{v-scaling}) and (\ref{Lt-scaling}),
it is possible to get a good approximation to almost every curve that results from the combination of parameters. To do this, 
all the parameters and the variable $E$ will be arguments of $T$,
\begin{equation}
T(E,G,V_0,Lt)\approx T(\alpha E,G-m,\frac{1}{k}V_0,\alpha Lt)^{k^2/\alpha^2 9^m},
\label{unified-scaling}
\end{equation}
for $G-m \geq 6$. All these approximations are quite useful to predict the transmission  
as function of the parameters, but it is necesary to test the correctness. So we calculate the root mean square of the difference between 
the target curve and the
scaled one, we call it $drms$. For parameter $G$ (scaling between generations of the potential, see figure \ref{gen-comparison}b) the $drms$ 
is $5.50\times 10^{-5}$; for $V_0$ (scaling in the energy of the main barrier, see figure \ref{H0Lt-comparison}b) the $drms$ is 
$2.05\times 10^{-4}$; at last for $Lt$ (scaling in the total length of the multibarrier structure, see figure \ref{H0Lt-comparison}d) the 
$drms$ is $1.43\times 10^{-3}$. From this it is quite clear that the better approximation is for the scaling between generations. 

This lead us to examinate a wider range of energy 
and check our scaling-laws values (see supplementary material).  
An analysis of the results, shows that the fit is excellent for small energies but tends to decrease in quality as the energy increases, as one 
should expect in the limiting case of very high energies. Also, the error tends to grow for small transmittance energies. However,
as seen in the supplementary material, if the error is computed for bigger generations, for example, for $G=7$ against scaled $G=8$, then
the error decreases in a dramatic way, as expected for a good scaling relation.  The drms for the Schroedinger case does not reduce 
as the generations grow, suggesting the absence of a fixed point. \newline

The numerical results presented above suggests that this type of potential indeed shows scaling properties, although 
this scaling is not found in a
single curve, but between curves corresponding to different parameters. This is a very interesting result because in fact the transmission coefficient
is an intricate combination of the wave functions, and even more, for a multibarrier potential it is the result of the product of several
transfer matrices. This suggests that the wave function have self-similar properties. That is the affirmation in the Gaggero-Pujals theorem
\cite{Gaggero-Sager-chap} for nonrelativistic electrons. Thus, it seems plausible that an extension of that theorem could be 
established for relativistic equations. Another important possibility is to study the nonperpedicular incidence, since Klein
tunneling is ruled out in this case \cite{katsnelson}.\newline

In conclusion, the self-similar potential proposed here show various kinds of scaling properties in the transmission curves. This scaling
properties gave the possibility to obtain, given a transmission curve and its set of parameters, an almost perfect approximation of the curve 
correspondig to certain transformation of the original parameters, see (\ref{unified-scaling}). It is also suggested that the scaling properties of 
the transmission curves are drastically affected by the mix of two geometric symmetries, for example, reflection and 
self-similarity, see \cite{Diaz-Guerrero2008,Gaggero-Sager-chap}. We believe that the proposed potential is 
ideal to study scaling properties in multibarrier systems, although our preliminary
results considering electrostatic self-similar barriers in grafene indicate also the presence of
a self-similar behavior.

Finally, we found that two ingredients are needed in order
to obtain a pure monofractal scaling of the transmittance, a {\it self-similar barrier and quantum relativistc equations}, since
the non-relativistic version does not display scaling. A possible heuristic explanation for such phenomena lies in the 
linear energy-momentum dispersion relation of the Dirac equation.

\end{document}